\documentclass[prx,twocolumn, superscriptaddress,
 showpacs,nofootinbib,
 longbibliography,
 aps,
 showkeys]{revtex4-2}

\usepackage{dsfont} % Typesetting the mathematical symbols for the natural numbers (N), whole numbers (Z), rational numbers (Q), real numbers (R) and complex numbers (C).
\usepackage{bm} % \bm to makfe bold fonts in math mode

\usepackage{bbold} %For identity
\usepackage{xcolor}
\usepackage{soul}
\usepackage{enumitem}

\usepackage{float}
\usepackage{physics}
\usepackage{graphicx,braket}
\usepackage{amssymb} 

\usepackage{hyperref}
\hypersetup{colorlinks=true,linkcolor=blue,citecolor=blue}
\usepackage{amsmath,amssymb,amsfonts}
\usepackage{wrapfig}
\usepackage[export]{adjustbox}
\usepackage{cleveref} % To reference mul000tiple equations
\usepackage{hhline}
\usepackage{siunitx}
\usepackage{booktabs}% http://ctan.org/pkg/booktabs
\def\bra#1{\mathinner{\langle{#1}|}} 
\def\ket#1{\mathinner{|{#1}\rangle}} %\newcommand{\braket}[2]{\langle #1|#2\rangle} 

\usepackage{enumitem}   

\usepackage{cleveref}% load last
\crefname{equation}{Eqs.}{Eqs.}
\Crefname{equation}{Equation}{Equations}% For beginning \Cref
\crefrangelabelformat{equation}{(#3#1#4--#5#2#6)}
\crefmultiformat{equation}{Eqs. (#2#1#3}{, #2#1#3)}{#2#1#3}{#2#1#3}
\Crefmultiformat{equation}{Equations (#2#1#3}{, #2#1#3)}{#2#1#3}{#2#1#3}

\usepackage{lipsum}

\begin{document}
\title{Entanglement of two optical emitters mediated by a terahertz channel}

\author{Yanis Le Fur}
\affiliation{Departamento de Física Téorica de la Materia Condensada and Condensed Matter Physics Center (IFIMAC),
Universidad Autónoma de Madrid, 28049 Madrid, Spain}
\affiliation{Institute of Fundamental Physics IFF-CSIC, Calle Serrano 113b, 28006, Madrid, Spain}

\author{Diego Martín-Cano}
\thanks{Corresponding author: \href{mailto:diego.martin.cano@uam.es}{diego.martin.cano@uam.es}.}
\affiliation{Departamento de Física Téorica de la Materia Condensada and Condensed Matter Physics Center (IFIMAC),
Universidad Autónoma de Madrid, 28049 Madrid, Spain}

\author{Carlos Sánchez Muñoz}
\thanks{Corresponding author: \href{mailto:carlos.sanchez@iff.csic.es}{carlos.sanchez@iff.csic.es}.}
\affiliation{Institute of Fundamental Physics IFF-CSIC, Calle Serrano 113b, 28006, Madrid, Spain}

\begin{abstract}
Quantum technologies in the terahertz (THz) require a coherent interface between addressable qubits and THz quantum channels---a capacity that so far, remains largely underdeveloped. 
Here, we propose and demonstrate the generation of steady-state entanglement between polar quantum emitters, mediated by THz photons.
We exploit strong visible-light driving of the emitters to create Rabi-split dressed eigenstates whose energy separation can be optically tuned into the THz regime. The polar nature of the emitters activates THz transitions within these eigenstates, allowing them to couple to a THz photonic mode that induces collective dissipative dynamics. A coherent driving and control of these effective THz emitters is achieved by using a sideband optical drive with detuning close to the THz transition frequency. The resulting interplay of collective dissipation and driving activates a mechanism to generate steady-state entanglement with high values of the concurrence ($\mathcal{C}>0.9$), attainable under experimentally feasible parameters. 
Crucially, both coherent manipulation and quantum state tomography are implemented entirely through optical means, avoiding direct THz control and detection. This establishes a hybrid visible–THz quantum interface in which a THz channel mediates qubit–qubit entanglement---a key operational requirement for THz quantum technologies---while remaining optically accessible.
\end{abstract}

% =============================================================================
% =============================================================================
\maketitle
% =============================================================================
% =============================================================================
\section{Introduction}
Quantum information platforms have reached a high degree of maturity in both microwave and visible regimes, providing a foundation for robust sensing and communication protocols. However, as these technologies tend to scale up, they encounter fundamental and relevant physical barriers. Superconducting circuits, while promisingly scalable, are strictly limited by the energy gap ($<1\,\text{THz}$)~\cite{tinkham2004,kerman2006,chremmos2010}, which imposes a hard ceiling on their operation frequency and requires millikelvin cryogenics. On the other hand, visible platforms operate at high frequencies and higher temperatures, but demands a nanometric fabrication precision to achieve the necessary small mode volumes and high quality factors, rendering large-scale integration an engineering challenge~\cite{vahala2003,koenderink2015,asano2017}. In this context, the terahertz (THz) regime emerges as a strategic compromise, offering potential for higher-temperature operation than superconductors, while requiring significantly less stringent fabrication tolerances than optical systems~\cite{todorov2024,buchwald2011}. The THz spectrum has already demonstrated advances in imaging~\cite{kawase2010,guerboukha2018}, spectroscopy~\cite{son2013,koch2023}, astronomy~\cite{li2025} and medical technologies~\cite{woodward2002,tonouchi2007,amini2021}. However, despite this large potential, THz quantum technologies have suffered a historical delay~\cite{todorov2024,tonouchi2007,siegel2002}. Unlike the visible or microwave domains, there  has been a critical lack of efficient coherent quantum emitters and interfaces in the THz spectrum ~\cite{kono2018,reiserer2013,blais2021,manetsch2025,iles-smith2025a,shishkov2025a}.
Despite growing efforts to bridge this gap through novel hardware~\cite{todorov2024,martin-cano2010,chen2021,chikkaraddy2023,chestnov2017,groiseau2024}, this limitation has continued to hamper the development of a fundamental milestone: the generation of entanglement mediated by a THz quantum channel~\cite{sherwin1999a,lee2011}. This capability would be an essential prerequisite for the realization of any viable THz based quantum network~\cite{narla2016,kimble2008,pezze2018}.

The interaction of a quantum system with its environment creates a significant challenge for the generation and stabilization of entanglement due to decoherence, rendering unitary protocols for entanglement generation unpractical in scenarios dominated by dissipation~\cite{divincenzo1997,srinivasa2024}. In those cases, an alternative strategy is to turn dissipation into a resource for stabilizing the target entangled state. Driven-dissipative protocols of generation of entanglement have been proposed~\cite{kraus2004,kraus2008,gonzalez-tudela2011,krauter2011,muschik2011,stannigel2012,pichler2015,didier2018,you2018,govia2022,agusti2022,harrington2022,agusti2023,lingenfelter2024,vivas-viana2024,agusti2025,chu2026} and 
realized experimentally in platforms such as superconducting circuits~\cite{irfan2025}, atomic ensembles~\cite{krauter2011} or trapped ions~\cite{lin2013}.

\begin{figure*}[hbt!]
    \centering
    \includegraphics[width=\textwidth]{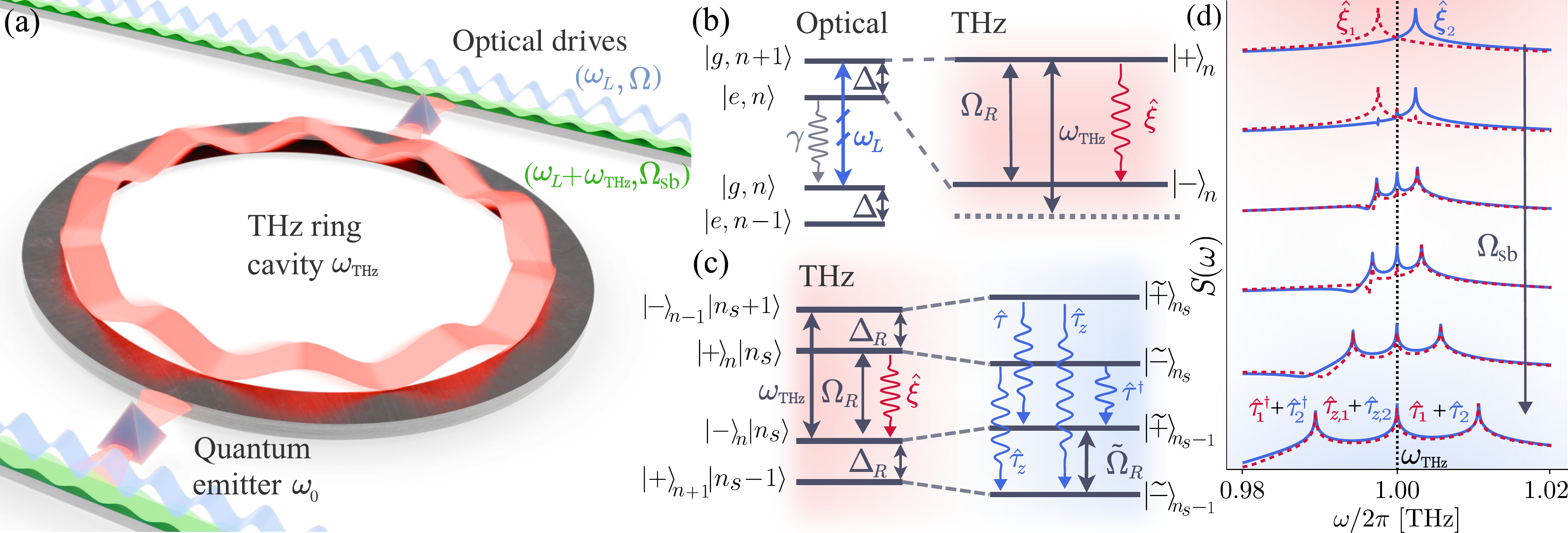}
    \caption{\textbf{Physical system and energy-level landscape}. (a) Schematic of the entanglement generation setup, consisting of two quantum emitters driven by carrier (blue) and sideband (green) optical fields, which are coupled to a shared THz channel (here represented by a ring waveguide or cavity). (b) Energy levels of an atom coupled to a carrier-laser. (Left) Bare system basis showing optical transitions in the laboratory frame; (Right) Dressed-state basis highlighting the emerging THz transitions within a Rabi doublet (red). (c) Energy levels of a dressed-atom coupled to a sideband laser: Dressed states split by THz frequencies couple to laser photons (left) developing a secondary dressed structure (right). (d) Emission spectra of individual emitters (red and blue) in the THz regime. (Upper) Without sideband drive ($\Omega_{\text{sb}} = 0$), each emitter shows a single peak, detuned from each other due to differences in dressed-state frequencies.  Activation of the sideband driving ($\Omega_{\text{sb}} \neq 0$) opens a secondary Mollow triplet; (Lower) The resulting Mollow triplets overlap, enabling the collective dissipation required for entanglement. Parameters: (d) $\Delta_1/2\pi = 871.6$ GHz, $\Delta_2/2\pi = 867.4$ GHz, $\Omega_1/2\pi = 499.8$ GHz, $\Omega_2/2\pi = 497.4$ GHz, $\omega_{\text{THz}}/2\pi = 1.0$ THz, $\gamma/2\pi = 39.79$ MHz, (d, Upper) $\Omega_{\text{sb},1} = \Omega_{\text{sb},2} = 0.0$, (d, Lower) $\Omega_{\text{sb},1}/2\pi = \Omega_{\text{sb},2}/2\pi = 10.3$ GHz. Here, the cavity-mediated qubit-qubit interaction is enforced to $J = 0$.}
     \label{fig:setup}
\end{figure*}

In this work, we establish a method to generate steady-state entanglement between two emitters mediated by a THz quantum channel [see Fig.~\ref{fig:setup}(a)]. While the transport and guiding of THz mediating modes have been extensively studied~\cite{martin-cano2010,vakil2011,huidobro2012}, the realization of efficient quantum emitters for the THz regime remains a significant challenge. Here, we use a recent proposal of quantum emitters in the THz domain~\cite{groiseau2024}, based on polar, optically-active emitters driven by a laser in the visible regime, which creates Rabi-split dressed states.
The permanent dipole moment---inherent in emitters with broken-inversion symmetry---activates radiative transitions between laser-dressed states of the same Rabi doublet, whose energy splitting can be optically tuned into the THz regime~\cite{savenko2012,shammah2014,deliberato2018,pompe2023, groiseau2024,groiseau2025}, see Fig~\ref{fig:setup}(b). These transitions can then couple to shared THz modes, enabling collective dissipative dynamics. The engineered dissipation is complemented with a coherent driving of the THz dressed-state transition by exciting with a secondary laser-field, detuned from the first field by approximately the THz mode frequency. 
The combination of collective dissipation and coherent driving activates a mechanism of dissipative stabilization of an entangled dark-state~\cite{stannigel2012,pichler2015,didier2018,you2018,govia2022,agusti2022,harrington2022,agusti2023,lingenfelter2024,vivas-viana2024,agusti2025}. We demonstrate with that protocol the generation of a high concurrence entangled state ($\mathcal{C} > 0.9$) under a feasible set of experimental parameters. To facilitate experimental verification, we propose a purely optical measurement reconstruction procedure based on quantum state tomography (QST)~\cite{james2001,altepeter2005} and linear inversion estimators, adapted from the classical shadow framework~\cite{huang2020,hu2022}. By incorporating result mitigation~\cite{srinivasan2022}, we show that the entangled state can be accurately reconstructed using current detection technologies~\cite{reiserer2013}, providing a clear pathway for the experimental realization and verification of THz-mediated quantum protocols. In summary, our proposed architecture introduces a novel hybrid interface where high-fidelity control and measurements are performed in the visible regime---where technology is more mature~\cite{chen2021,reiserer2013,williams2007}---while long-range quantum correlations are mediated by the engineered THz environment.

\section{Model}
\label{sec:model}

We consider a pair of non-identical polar emitters modeled as two-level systems (TLS) with ground states $\ket{g_i}$,  excited states $\ket{e_i}$ and a transition frequency  $\omega_i$ in the visible regime. These emitters are driven by two external optical fields: a carrier field $\mathbf{E}_L$ and the sideband field $\mathbf{E}_\text{sb}$ respectively at visible frequencies $\omega_L$ and $\omega_L+\omega_{\text{THz}}$. Both emitters are coupled to a THz channel that mediates their mutual interactions. We model this channel as a lossy single-mode resonator centered at frequency $\omega_{\text{THz}}$ operating in the bad-cavity limit~\cite{walls2008}. In this regime, the large decay rate prevents the build-up of a large intra-cavity photon population and enforces irreversible dissipative dynamics. Consequently, this cavity mode can be adiabatically eliminated, rendering the dynamics equivalent to those of a linear waveguide and thus providing a general description of a dissipative THz quantum channel~\cite{gonzalez-tudela2017}. While our model is formulated in a general way,  Fig.~\ref{fig:setup}(a) depicts a possible implementation of this THz channel as a ring resonator coupling to  both quantum emitters~\cite{chremmos2010}. 

The total Hamiltonian of the system is given by $\hat{H} = \hat{H}_{1}+\hat{H}_{2}+\omega_{\text{THz}}\hat{a}^\dagger{}\hat{a}$ (with $\hat{a}$ being the annihilation operator of the cavity mode), where the dynamics of the $i$-th qubit is governed by: $\hat{H}_{i} = \frac{\omega_i}{2} \hat\sigma_{z,i}  + \mathbf{\hat{d}}_i \cdot \mathbf{E}_c (\hat{a} + \hat{a}^\dagger) + \mathbf{\hat{d}}_i \cdot \mathbf{E}_{L,i} \cos(\omega_L t) +  \mathbf{\hat{d}}_i \cdot \mathbf{E}_{\text{sb},i}\cos[(\omega_L+\omega_{\text{THz}})t]$. The dipole operator $\mathbf{\hat{d}}_i = \mathbf{d}_{ee,i}(1+\hat{\sigma}_{z,i})/2+\mathbf{d}_{ge,i}(\hat{\sigma}_{i}^\dagger+\hat{\sigma}_{i})$ includes a permanent dipole component $\mathbf{d}_{ee}$, which arises from charge distribution asymmetries~\cite{chestnov2017}. Here, $\hat \sigma_{i}$ are the Pauli lowering operators acting on the $i$-th emitter, $\hat\sigma_{z,i}  = 2\hat\sigma_{i}^\dagger\hat\sigma_{i}-\mathbb{I}$, and $\hat\sigma_{x,i} = \hat\sigma_{i}+\hat\sigma_{i}^\dagger{}$ are the Pauli basis operators. Crucially, this permanent dipole moment enables resonant interaction between the THz cavity and the dressed emitter~\cite{groiseau2024}. To see this, we transform the emitters to a frame rotating at the carrier frequency $\omega_L$ and apply the rotating wave approximation (RWA) to the terms rotating at visible frequencies, obtaining the Hamiltonian {$\hat H(t) = \hat H_\text{d} + \hat H_\text{sb}(t)$, given by
\begin{align}
\hat{H}_{\text d} &=
\omega_{\text{THz}}\hat{a}^\dagger \hat{a} + \frac{\Delta_1}{2}\hat\sigma_{z,1} + \frac{\Delta_2}{2}\hat\sigma_{z,2} \notag
\\
&
+\sum_{i=1,2}
\left[
 \frac{\Omega_i}{2}\hat\sigma_{x,i}+
\chi_i (1+\hat\sigma_{z,i})(\hat a + \hat a^\dagger)
\right],
\label{eq:bare_hamiltonian} \\
\hat H_{\text{sb}}(t) &=
\frac{\Omega_{\text{sb},i}}{2}
\left(\hat\sigma_{i} e^{i\omega_{\text{THz}} t}
+ \text{H.c.}\right),
\label{eq:sideband_hamiltonian}
\end{align}
where $\Delta_i = \omega_i-\omega_L$ is the laser detuning, $\chi_i = \mathbf{d}_{ee,i}\cdot{}\mathbf{E}_{c,i}/2$ is the cavity-emitter coupling and  $\Omega_i = \mathbf{d}_{ge,i}\cdot{}\mathbf{E}_{L,i}$ and $\Omega_{\text{sb},i} = \mathbf{d}_{ge,i}\cdot{}\mathbf{E}_{\text{sb},i}$ are the Rabi frequencies for the carrier and sideband drives, respectively~\cite{scully1997,chestnov2017}. To describe the open-system dynamics, we account for the single-emitter optical emission at rate $\gamma$ and the cavity photon losses at rate $\kappa$. Given the presence of counter-rotating terms in $\hat H_\text{d}$ in Eq.~\eqref{eq:bare_hamiltonian} with rates $\chi_i$ and cavity photon losses at rate $\kappa$ that may be comparable to the cavity frequency $\omega_\text{THz}$, a standard Lindblad decay term with operator $\hat a$ could lead to non-physical processes~\cite{beaudoin2011,settineri2018} (this is not the case for the decay of the emitters, since these oscillate at visible frequencies). We therefore describe the cavity-environment interaction using the operator $\hat{X}^+ = \sum_{j,k>j} \sqrt{\omega_{jk}/\omega_{\text{THz}}}\bra{j}(\hat{a}+\hat{a}^\dagger{})\ket{k}\ket{j}\bra{k}$ that encompass all the positive-frequency transition of $(\hat{a}+\hat{a}^\dagger)$ where  $\ket{k}$ is the k-eigenstate of $\hat H_\text d$  and $\omega_{jk} = \omega_j-\omega_k$ the difference between eigen-energies~\cite{breuer2007,mercurio2023} (the sideband-drive Hamiltonian $\hat H_\text{sb}$ will be considered a small perturbation and not included in this diagonalization procedure). The complete dynamics of the system are governed by the master equation~\cite{carmichael1993} 
% ==== MASTER EQUATION =====
$\partial_t\hat{\rho} = -i[\hat H(t),\hat{\rho}]+\gamma\mathcal{D}[\hat{\sigma}_{1}]+\gamma\mathcal{D}[\hat{\sigma}_{2}]+\kappa\mathcal{D}[{\hat{X}^+}]$
where the Lindblad super-operator is defined as $\mathcal{D}[\hat{O}] = \hat{O}\hat{\rho}\hat{O}^\dagger{}-\frac{1}{2}\left\{\hat{O}^\dagger{}\hat{O},\hat{\rho}\right\}$. 

\subsection{Dressed qubit basis}
\label{sec:dressed}
The carrier drive dresses the bare emitter states, creating a new eigenbasis of the emitter-laser subsystem split by the Rabi frequency $\Omega_{R,i} = \sqrt{\Delta_i^2+\Omega_i^2}$ [see Fig.~\ref{fig:setup}(b)]~\cite{munoz2017}. These dressed eigenstates are defined by $\ket{+_i}= s_i\ket{g_i}+ c_i\ket{e_i}$ and $\ket{-_i}= c_i\ket{g_i}- s_i\ket{e_i}$ where $s_i \equiv \sin \theta_i$ and $c_i \equiv \cos \theta_i$ with the mixing angle $\theta_i \equiv \arctan\left(\frac{\Omega_{R,i}-\Delta_i}{\Omega_i}\right)$~\cite{chestnov2017,groiseau2024}. The Pauli and lowering operators $\hat\sigma_{z,i}$, $\hat\sigma_i$  can be expressed in terms of the equivalent operators in the dressed-state basis $\hat\xi_{z,i}$, $\hat\xi_{x,i}$, and $\hat\xi_i$ as: $\hat\sigma_{z,i} = (c_i^2-s_i^2)\hat\xi_{z,i}-2c_is_i\hat\xi_{x,i}$ and $\hat\sigma_{i}  = c_i^2\hat\xi_{i}-s_i^2\hat\xi_{i}^\dagger+c_is_i\hat\xi_{z,i}$. By tuning the laser parameters $\Omega$ and $\omega_L$, the splitting $\Omega_{R,i}$ can be brought close to resonance with the THz cavity frequency $\omega_{\text{THz}}$, enabling the emission of THz photons via transitions between the Rabi doublet states [see Fig.~\ref{fig:setup}(c)]~\cite{groiseau2024}. To describe explicitly the interactions between these dressed states and the THz cavity, we can transform the Hamiltonian into the dressed basis:

\begin{equation}
    \begin{split}
        \hat{H} = &\omega_{\text{THz}}\hat{a}^\dagger{}\hat{a}+\frac{\Omega_{R,1}}{2}\hat\xi_{z,1}+\frac{\Omega_{R,2}}{2}\hat\xi_{z,2}\\
        &+\sum_{i=1,2}
        \left\{\chi_i\left(\hat{a}+\hat{a}^\dagger{}\right)\left[1+\left(c_i^2-s_i^2\right)\hat\xi_{z,i}-2c_is_i\hat\xi_{x,i}\right] 
        %add vphantom just to adjust size, don't freak out Yanis
        \vphantom{ \frac{\Omega_{\text{sb},i}}{2}}
        \right. \\
    & \left. + \frac{\Omega_{\text{sb},i}}{2}\left[\left(2c_is_i\hat\xi_{z,i}+c_i^2\hat\xi_{i}-s_i^2\hat\xi_{i}^\dagger\right)e^{i\omega_{\text{THz}}{}t}+\text{H.c}\right] \right\} .
    \label{eq:dressed_hamiltonian}
    \end{split}
\end{equation}
The presence of permanent dipoles ($\mathbf{d}_{ee}$) in the bare basis leads to a longitudinal coupling term in the dressed basis proportional to $\chi_i\hat\xi_{z,i}(\hat{a}+\hat{a}^\dagger)$. A standard rotating wave approximation (RWA) would neglect this term. However, this longitudinal interaction induce a state displacement of the cavity field, resulting in a renormalization of the qubit frequencies. This effect needs to be taken into account, since the generation of entanglement is critically sensitive to such frequency shifts~\cite{vivas-viana2024}.  To capture these shifts accurately, we employ a polaron transformation~\cite{Lang1963,debernardis2025} prior to performing the RWA, a procedure known as the Generalized Rotating Wave Approximation (GRWA)~\cite{twyeffortirish2007,zhang2015,montano2023}.
Assuming the weak coupling regime relative to the cavity frequency ($\omega_{\text{THz}}\gg\chi$) and weak sideband driving ($\omega_{\text{THz}}\gg\Omega_{\text{sb},i}$), we move all modes to a frame rotating at $\omega_\text{THz}$ and apply the  GRWA to get the effective Hamiltonian $\hat H = \hat H_q + \hat H_c$, where (see App.~\ref{app:polaron} for the full derivation):
\begin{eqnarray}
    \hat{H}_q &=& \sum_{i=1,2}\left(\frac{\Delta_{R,i}}{2}\hat{\xi}_{z,i} + c_i^2\frac{\Omega_{\text{sb},i}}{2}\hat{\xi}_{x,i}\right)-J\hat{\xi}_{z,1}\hat{\xi}_{z,2},
     \label{eq:ad_hamiltonian} \\
    \hat{H}_c &=& \sum_i-2\chi{}c_is_i\left(\hat{a}^\dagger{}\hat{\xi}_{i}+\hat{a}\hat{\xi}_{i}^\dagger\right).
     \label{eq:Hc}
\end{eqnarray}
Here, the polaron transformation results in an effective Lamb-shift in the dressed-emitter detunings, $\Delta_{R,i} = \Omega_{R,i}-\omega_{\text{THz}}-\Lambda_i $, with $\Lambda_i \equiv 8\chi^2(c_i^2-s_i^2)/\omega_{\text{THz}}$, and an effective qubit-qubit interaction with rate $J \equiv 2\chi^2(c_1^2-s_1^2)(c_2^2-s_2^2)/\omega_{\text{THz}}$.
We can now solve a time-independent master equation given by
\begin{equation}
    \partial_t\hat{\rho} = -i[\hat H_q + \hat H_c,\hat{\rho}]+\gamma\mathcal{D}[\hat{\sigma}_{1}]+\gamma\mathcal{D}[\hat{\sigma}_{2}]+\kappa\mathcal{D}[{\hat{X}^+}].
    \label{eq:me-grwa}
\end{equation}

For further analytical understanding, we will focus on the bad-cavity limit ($\kappa \gg c_i s_i\chi$) where the cavity decay is the dominant timescale.  Adiabatically eliminating the cavity mode and considering its linewidth broad enough to not resolve dressed-qubit splittings--- $\kappa\gg\Delta_{R,i}$~\cite{vivas-viana2024,navarrete-benlloch2022}---leads to the effective master equation given by
        \begin{equation}
\partial_t\hat{\rho} = -i[\hat{H}_q,\hat{\rho}]+\sum_i\gamma\mathcal{D}[\hat{\sigma}_{i}]+\mathcal D[\hat L],
             \label{eq:master_eq_dressed}
        \end{equation}
where the effect of the cavity is described by the collective jump operator $\hat L \equiv   \sqrt{\Gamma_1}\hat{\xi}_{1}+\sqrt{\Gamma_2}\hat{\xi}_{2} $, with $\Gamma_i = \frac{4(-2c_is_i\chi)^2}{\kappa}$ the Purcell decay rate. In this limit, the cavity acts as a passive bus via the collective Purcell effect which, as we discuss below, mediates entanglement generation~\cite{gonzalez-tudela2011,pichler2015,vivas-viana2024}.

\subsection{Doubly-dressed qubit basis}
\label{sec:model_doubly-dressed}

Due to the introduction of the sideband field, the Hamiltonian $\hat H_q$ in Eq.~\eqref{eq:ad_hamiltonian} corresponds to dressed emitters that are themselves driven with Rabi frequencies $c_i^2\Omega_{\text{sb},i}$, resulting in a second layer of hybridization [see Fig.~\ref{fig:setup}(c)]. Just as the carrier field split the bare atomic levels, this sideband drive splits the dressed states into doubly-dressed states $\ket{\tilde{+}},\ket{\tilde{-}}$~\cite{yan2001} with an effective transition frequency $\tilde{\Omega}_{R,i} = \sqrt{\Delta_{R,i}^2+c_i^2\Omega_{\text{sb},i}^2}$. Spectrally, this secondary splitting manifest as a secondary Mollow triplet~\cite{mollow1969}, appearing both in the THz domain, centered around the THz emission peak at $\sim\omega_{\text{THz}}$ [see  Fig.~\ref{fig:setup}(d)], as well as in the visible domain, centered around the higher-energy optical sideband of the primary Mollow triplet created by the carrier drive [see bottom panel of  Fig.~\ref{fig:variation_drive}(a)]. Notice that, while the primary Mollow sidebands are split from the main peak by THz frequencies, the secondary splittings need to be much lower (e.g. GHz frequencies) due to the GRWA validity condition, which requires $\Omega_{\text{sb},i}\ll\omega_\text{THz}$. We define a doubly dressed basis as $\ket{\tilde +_i}= \tilde s_i\ket{-_i}+ \tilde c_i\ket{+_i}$ and $\ket{\tilde-_i}= \tilde c_i\ket{-_i}- \tilde s_i\ket{+_i}$, where  the mixing factors are $\tilde{s}_i = \sin\tilde{\theta}_i$ and $\tilde{c}_i = \cos\tilde{\theta}_i$ with the doubly-dressed mixing angle $\tilde{\theta}_i = \arctan\left(\frac{\tilde{\Omega}_{R,i}-\Delta_{R,i}}{c_i ^2 \Omega_{\text{sb},i}}\right)$. By defining the Pauli and lowering operators in the doubly-dressed basis ($\hat\tau_{z,i}$, $\hat\tau_{x,i}$, $\hat\tau_{i}$), and transforming the Hamiltonian (\ref{eq:ad_hamiltonian}) to this basis, we get
\begin{equation}
        \hat{H}_q = \sum_i \frac{\tilde{\Omega}_{R,i}}{2}\hat\tau_{z,i} + {J}\prod_i \left[\left(\tilde{c}_i^2-\tilde{s}_i^2\right)\hat\tau_{z,i}-2\tilde{c}_i\tilde{s}_i\hat\tau_{x,i}\right].
        \label{eq:doubly_dressed_hamiltonian}
    \end{equation}

In the strong-driving limit ($\tilde{\Omega}_{R,i}\gg\Gamma_i$), a RWA allows us to split the original collective dissipation operator $\hat L$  into three distinct Lindblad terms corresponding to the three peaks of the secondary Mollow triplet~\cite{govia2022}:
\begin{equation}
        \begin{split}
            \hat{L}_+&=\sqrt{\Gamma_1}\tilde{c}_1^2\hat\tau_{2}+\sqrt{\Gamma_2}\tilde{c}_2^2\hat\tau_{1},\\
            \hat{L}_-&=\sqrt{\Gamma_1}\tilde{s}_1^2\hat\tau_{1}^\dagger{}+\sqrt{\Gamma_2}\tilde{s}_2^2\hat\tau_{2}^\dagger{},\\
            \hat{L}_z &= \sqrt{\Gamma_1}\tilde{c}_1\tilde{s}_1\hat\tau_{z,1}+\sqrt{\Gamma_2}\tilde{c}_2\tilde{s}_2\hat\tau_{z,2}.
        \end{split}
        \label{eq:sideband_jumps}
    \end{equation}

\subsection{Conditions for entanglement generation}
\subsubsection{General conditions}
This dissipative structure enables the stabilization of a dark state~\cite{irfan2025,govia2022} given by
\begin{equation}
    \ket{D} = \tilde{c}_1^2\ket{\tilde{+}\tilde{-}}-\tilde{c}_2^2\ket{\tilde{-}\tilde{+}}.
    \label{eq:dark_state}
\end{equation}
For $\ket{D}$ to be a dark state, it must simultaneously be the eigenstate of the Hamiltonian in Eq.~\eqref{eq:doubly_dressed_hamiltonian} and lie in the kernel of the jump operators in Eq.~\eqref{eq:sideband_jumps}. 
As we elaborate in Appendix~\ref{app:doubly-dressed}, these two requirements translate respectively into the following parametric conditions for the stabilization of a pure dark state:  

\begin{enumerate}[label=\textbullet\ Condition \arabic*:,       leftmargin=0pt,
    labelsep=0.5em,
    itemindent=!,
    align=left]
    \item \emph{Symmetric primary Mollow sideband}.
    \begin{equation}
    \sqrt{\Gamma_1}\tilde{c}_1\tilde{s}_1 = \sqrt{\Gamma_2}\tilde{c}_2\tilde{s}_2.
    \end{equation}
    This condition ensures $|D\rangle$ lies in the kernel of the jump operators. % 
    In general terms, this implies an equal rate multiplying the $\tau_{z,1/2}$ in Eq.~\eqref{eq:sideband_jumps}, i.e., equal linewidth for the central secondary-Mollow peak of each individual emitter. More particularly, if $\Gamma_1\sim\Gamma_2$, this condition is met by placing the higher-energy primary-Mollow sidebands of the two emitters symmetrically about the sideband drive frequency, as we elaborate below.
    \item \emph{Overlapping secondary-Mollow sidebands}.
    \begin{equation}\tilde\Omega_{R,2} = \tilde\Omega_{R,1}+J\sin2\tilde\theta_1\sin2\tilde\theta_2\left(\frac{\sqrt{\gamma_1}\tilde{c}_1^2}{\sqrt\gamma_2\tilde{c}_2^2}-\frac{\sqrt{\gamma_2}\tilde{c}_2^2}{\sqrt\gamma_1\tilde{c}_1^2}\right).
    \label{eq:condition-2}
    \end{equation}
\end{enumerate}
This condition ensures that $|D\rangle$ is an eigenstate of $\hat H_q$. Since $J$ is a small effective polaron coupling, this approximately amounts to $\tilde\Omega_{R,2} \approx \tilde\Omega_{R,1}$, i.e., the two secondary Mollow triplets share the same Rabi frequency and are therefore spectrally overlapping.
\subsubsection{Symmetric cavity coupling}
\label{sec:conditions-symmetric}
In the following, we will assume a symmetric coupling to the cavity ($\Gamma_1 = \Gamma_2$), which is optimal for entanglement generation and results in a simpler, more transparent set of conditions for the stabilization of a dark state. Adding this extra condition, the set of requirements for the tunable optical parameters can be written as:
\begin{enumerate}[
    label=\textbullet\ Condition \arabic*:,
    start=0,
    leftmargin=0pt,
    labelsep=0.5em,
    itemindent=!,
    align=left]
   % Condition 0
   % =========
    \item \emph{Symmetric cavity coupling}:
 \begin{equation}
       \frac{\Omega_1}{\Delta_1} = \frac{\Omega_2}{\Delta_2}
       \label{eq:condition_0}
 \end{equation}   
This condition, which assumes $\chi_1 = \chi_2$, directly implies $c_1 s_1 = c_2 s_2$ and thus $\Gamma_1 = \Gamma_2$. 
   % Condition 1
   % =========
    \item \emph{Symmetric primary-Mollow sidebands}:
    \begin{equation}
        \Delta_{R,1} = -\Delta_{R,2} \equiv \Delta/2.
    \end{equation}
This is equivalent to the the complementary angle condition ($\tilde\theta_1 = \pi/2-\tilde\theta_2 \equiv \tilde \theta$).
   This condition can be met by fine-tuning $(\Omega_i,\Delta_i)$, maintaining the constraint of Condition 0.
   % Condition 2 
   % =========
    \item \emph{Overlapping secondary-Mollow sidebands}. \begin{equation}
        \tilde\Omega_{R,2} = \tilde\Omega_{R,1}+ \epsilon \equiv \tilde\Omega_R,
    \end{equation}
\end{enumerate}
with $\epsilon\equiv 4J\cos 2\tilde\theta$.

\begin{figure*}[hbt!]
   
    \centering
    \includegraphics[width=\textwidth]{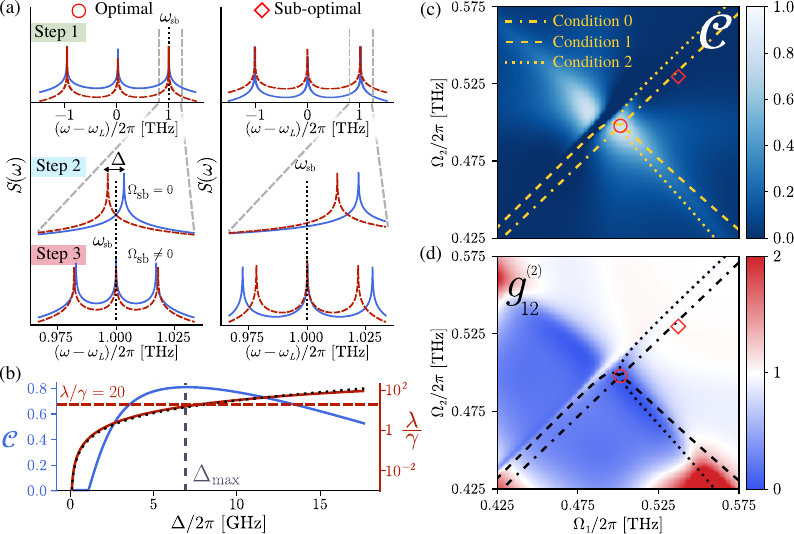}
    \caption{
    % (a)
    \textbf{Optimizing entanglement via optical measurements}
      (a) Emission spectrum in the visible regime for each individual emitter, centered at the laser frequency $\omega_L$ for the case of both optimal and sub-optimal spectral alignments. 
      %Step 1
      (Top) Spectra in the absence of sideband drives ($\Omega_\text{sb} = 0$) obtained in Step 1, resulting in Mollow triplets with THz sideband splittings.
      % Step 2
      (Middle) Zoom of the higher-energy Mollow sideband peaks, aligned symmetrically about $\omega_\text{sb}$ at the end of Step 2. 
      % Step 3
      (Bottom) Formation of secondary Mollow spectra upon activation of the sideband drive $\Omega_\text{sb}$, which should be spectrally overlapping at the end of Step 3. (b) Concurrence $\mathcal{C}$ (blue, solid) and Liouvillian gap  $\lambda$---computed numerically (solid, red) and analytically from Eq.~\eqref{eq:liouvillian_gap} (dotted black)---versus the detuning $\Delta$ (with $\tilde\Omega_R$ fixed). (c) Stationary concurrence $\mathcal{C}$ and (d) the associated normalized second-order cross-correlation function $g_{12}^{(2)}$ versus carrier drive amplitudes ($\Omega_1$,$\Omega_2$). Conditions 0 to 2 define three curves in the ($\Omega_1$,$\Omega_2$) plane, with their intersection (red circle) defining the optimal operating point. The red diamond denotes a low-concurrence reference point.  Parameters: (a, red circle) $\Omega_1/2\pi= 499.7$ GHz, $\Omega_2/2\pi = 496.3$ GHz (a, red diamond) $\Omega_1/2\pi= 537.3$ GHz, $\Omega_2/2\pi = 529.7$ GHz. (b) $\tilde\Omega_{R}/2\pi = 16.0 $ GHz, with the rest of parameters derived from $(\tilde\Omega_{R},\tilde\theta)$ as discussed in Appendix~\ref{sec:appendix_optim_params}. (a,c,d) $\Delta_1/2\pi = 874.9$ GHz, $\Delta_2/2\pi = 868.9$ GHz, $\Omega_{\text{sb},1}/2\pi = 16.7 $ GHz, $\Omega_{\text{sb},2}/2\pi = 17.5 $ GHz. (a-d) $\gamma/2\pi =39.79$ MHz, $\kappa/2\pi = 59.6$ GHz, $\chi/2\pi = 24.4$ GHz and $\omega_{\text{THz}}/2\pi = 1.0$ THz.
    }
    \label{fig:variation_drive}
\end{figure*}
To summarize, these conditions mean that (1) the original Mollow sidebands from the two emitters should be centered at either sides of the sideband-drive frequency and split by $\Delta$, and (2) that the secondary Mollow triplets emerging from the sideband drive should almost overlap, sharing the same Rabi frequency $\tilde\Omega_R$.
\subsubsection{Tradeoff between entanglement and stabilization rate}
The conditions above do not fix the values of $\tilde \Omega_R$ and $\Delta$, which remain free tunable parameters whose ratio determines the secondary dressing angle $\tilde\theta$.  Crucially, there is a non-monotonic dependence of the degree of entanglement on the angle $\tilde\theta$, leading to a nontrivial optimal value that maximizes entanglement. This arises from a fundamental trade-off between the entangled nature of the dark state and the time required for its stabilization~\cite{brown2022,pocklington2024,irfan2024}, which is set by the Liouvillian gap and dependent on $\tilde\theta$. If the conditions discussed above are met, the Liouvillian gap is given by (see App.\ref{app:Lgap}): 
\begin{equation}
    \lambda \approx \frac{4\Gamma_1}{3}\cot^2(2\tilde{\theta}).
    \label{eq:liouvillian_gap}
\end{equation}
The condition for the dark state $|D\rangle$ in Eq.~\eqref{eq:dark_state} to be a maximally entangled Bell state is reached in the limit $\tilde{\theta}\rightarrow \pi /4$, i.e., when $\Delta \rightarrow 0$. However, the system must be able to reach and maintain the state faster than  decoherence destroys it, meaning that the Liouvillian gap must be significantly higher than the rate of spontaneous emission of the emitters ($\lambda\gg\gamma$) or of any other source of decoherence. Crucially, Eq.~\eqref{eq:liouvillian_gap} shows that these two requirements are inversely linked. While the entanglement is maximized when $\tilde\theta$ tends toward $\pi/4$, the Liouvillian gap is simultaneously closed. This competition defines an optimization problem where one must find the optimal angle to maximize the steady-state entanglement. This angle can be roughly estimated to correspond to the point where $\lambda \sim 20\gamma$, a point below which the Liouvillian gap is too small to compete against spontaneous emission.

\subsubsection{Practical implementation strategy}
We now translate the insights and conditions discussed above into a fully optical experimental strategy for preparing and optimizing the entangled steady-state. This strategy consists of a sequence of steps based solely on tuning optical parameters guided by spectral measurements, as detailed below and illustrated in Fig.~\ref{fig:variation_drive}(a):
\begin{enumerate}[label=\(\diamond\) Step \arabic*:,     leftmargin=0pt,
    labelsep=0.5em,
    itemindent=!,
    align=left]
    % === STEP 1 =======
    \item Initial dressing: Drive the bare emitters using only the carrier lasers. Ensure symmetric coupling to the cavity ($\Gamma_1= \Gamma_2$) by setting (assuming $\chi_1 = \chi_2$):
\begin{align}
        \Omega_1 & = \Omega_2,\\
        \Delta_1 & = \Delta_2.        
    \end{align}
    This choice satisfies Eq.~\eqref{eq:condition_0}, thereby fulfilling Condition 0.
    %
    % === STEP 2 =======
    \item Spectral tuning to meet Condition 1. From the starting values set in Step 1, fine-tune the carrier parameters ($\Omega_i$, $\Delta_i$), while maintaining the constraint in Eq.~\eqref{eq:condition_0}, such that the higher-energy Mollow sideband peaks are positioned symmetrically around the sideband laser frequency ($\omega_\text{sb} = \omega_\text L + \omega_\text{THz}$). 
    If only the drive amplitudes $\Omega_i$ can be tuned experimentally to satisfy Condition~1, the protocol remains sufficiently robust to experimental misalignments such that small deviations from the equality imposed by Condition 0 are not expected to significantly affect the outcome.
    % === STEP 3 =======
    \item Sideband activation to meet Condition 2. Turn on the sideband drives ($\Omega_\text{sb,i}$), generating
     secondary Mollow spectra for each emitter. The values of $\Omega_\text{sb,i}$ should be chosen so that these spectra overlap. Notice that $J$ coupling may require non-symmetric values of sideband drive, i.e. $\Omega_{\text{sb},1}\neq \Omega_{\text{sb},2}$ due to the $\epsilon$ in Condition 2.
     % === STEP 4=======
    \item Fine tuning. At this stage, $\tilde \Omega_R$ and $\Delta$ remain free tunable parameters whose ratio determines the secondary dressing angle $\tilde\theta$.  
    As discussed above, the entanglement is maximum for a nontrivial value of $\tilde\theta$. In this step, we identify this optimum through fine tuning of either $\Delta$---which may imply readjusting the ($\Omega_i$, $\Delta_i$) values set in Step 2 while maintaining the same constraints---, or $\Omega_{\text{sb},i}$.
\end{enumerate}

\section{Results}

\subsection{Entanglement quantification}
To quantify the generated entanglement, we use the concurrence $\mathcal C$ \cite{hill1997,wootters1998}, calculated from the steady-state density matrix of the GRWA master equation in Eq.~\eqref{eq:me-grwa}. 
The discussed non-monotonic dependence of $\mathcal C$ on $\tilde \theta$ is shown in Fig.~\ref{fig:variation_drive}(b), where $\tilde\theta$ is modified done by varying the value of $\Delta$ (which can be fixed at Step 2). 

 Another example of entanglement tuning is shown in Fig.~\ref{fig:variation_drive}(c), depicting the concurrence versus the two carrier drive amplitudes $(\Omega_{1},\Omega_{2})$ with all other parameters fixed. Conditions 0, 1 and 2 define three curves in the $(\Omega_{1},\Omega_{2})$ space, whose crossing sets the point where all the conditions are satisfied and entanglement is maximized, as confirmed by the numerical calculation of the concurrence $\mathcal C$ shown as a heatmap. 

 While the verification the entanglement via a full tomographic reconstruction of the density matrix of the emitters is possible in practice---as we show in following sections---, the steady-state second-order cross-correlation function $g_{1,2}^{(2)} = \langle \hat\sigma_{1}^\dagger\hat\sigma_{2}^\dagger\hat\sigma_1\hat\sigma_2\rangle /
 \langle \hat\sigma_1^\dagger\hat\sigma_1 \rangle \langle \hat\sigma_2^\dagger \hat\sigma_2 \rangle$, which quantifies the probability of simultaneous emission of photons from both quantum emitters, serves as a quick, simple 
 entanglement witness.  Since maximum entanglement is achieved by a dark state confined into the single-excitation manifold~\cite{Dicke1954,Scully2015}, the concurrence is anticorrelated with $g_{12}^{(2)}$, as we show in Fig.~\ref{fig:variation_drive}(d). This implies that minimizing this readily observable quantity provides a route towards the parametric regions of optimal entanglement.

\begin{figure}[h!]
    \centering
    \includegraphics[width=0.95\linewidth]{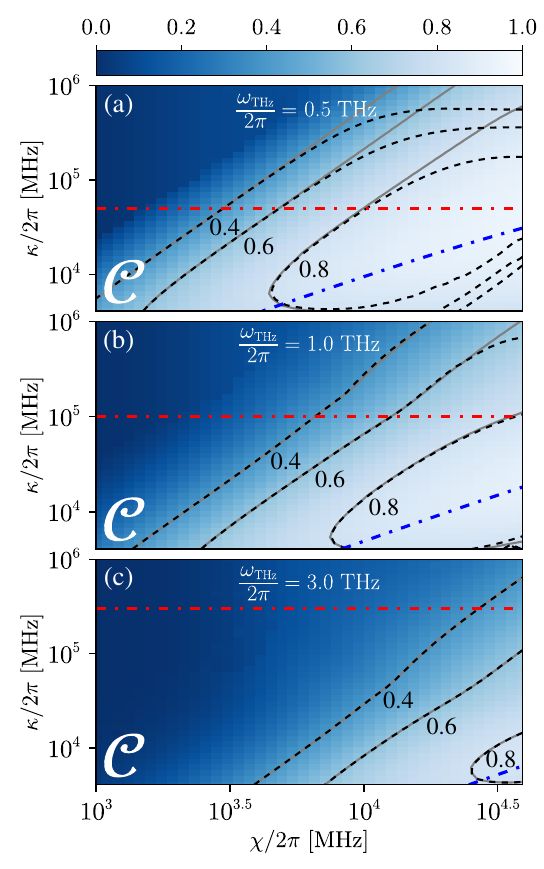}
    \caption{Maximal stationary concurrence $\mathcal{C}$ versus cavity loss rate $\kappa$ and emitter-cavity coupling $\chi$, found by optimizing the doubly dressed angle $\tilde\theta$ and the doubly dressed frequency $\tilde\Omega_{R}$ at every $(\kappa,\chi)$ pair (the rest of parameters are fixed by Conditions 0-2). Calculations were performed using the GRWA framework Eq.~\eqref{eq:ad_hamiltonian} and ~\eqref{eq:Hc} and subsequently verified against the full model from (\ref{eq:bare_hamiltonian}). The dash-dotted lines represent the different condition we have imposed for our analytical findings: the adiabatic elimination of the cavity $\kappa\geq2c_1s_1\chi$ (blue) and the RWA on the dissipation of the cavity $\omega_{\text{THz}}\geq 10\kappa$ (red). Panels correspond to cavity frequencies of (a) $\omega_{\text{THz}}/2\pi = 0.5$~THz, (b) $\omega_{\text{THz}}/2\pi = 1.0$~THz and (c) $\omega_{\text{THz}}/2\pi = 3.0$~THz. Parameters: $\gamma/2\pi = 39.79$ MHz. The optimal values found for $\tilde\Omega_{R}$ and $\tilde{\theta}$ at each point in these maps are respectively presented in Appendix~\ref{sec:appendix_optim_params} (Fig.~\ref{fig:optimal_omega_r} and Fig.~\ref{fig:optimal_theta_1}).}
    \label{fig:optimization_map}
\end{figure}

To establish the practical limits of our protocol, we explore a wide set of THz cavity parameters $(\chi,\kappa)$ (see Fig.~\ref{fig:optimization_map}).  
We assume a fixed cavity frequency $\omega_\text{THz}$ for each map of Fig.~\ref{fig:optimization_map} and set the emitter decay rate to $\gamma/2\pi \approx{}39.79$ MHz~\cite{Esmann2024}. To make this optimization computationally feasible, we drastically reduce the initial six-dimensional parameter space --- spanning the carrier and sideband drives $(\Omega_i, \Delta_i, \Omega_{\text{sb},i})$ --- down to just two free parameters chosen to be $(\tilde\Omega_{R,1}, \tilde\theta_1)$ by applying Conditions 0-2 and capping the maximum allowed $\Omega_i < \Omega_\text{max} = 0.5\,\text{THz}$ (see Appendix~\ref{sec:appendix_optim_params}).

For each cavity configuration, we numerically maximize the concurrence over  $(\tilde\Omega_{R,1},\tilde\theta_1)$ using the steady-state solution of the GRWA master equation Eq.~\eqref{eq:me-grwa}. Once the optimal parameters have been found this way, we verify the validity of the GRWA by using those optimal parameters in a master equation governed by  the full time-dependent Hamiltonian \eqref{eq:bare_hamiltonian}, obtaining an effective steady state by averaging the small oscillations of the density matrix in the long time limit.
Our simulations predict a robust entanglement generation across the THz spectrum. Specifically, we achieve maximum steady-state concurrences of $\mathcal{C}\approx 0.91,0.90$ and $0.83$ for cavity frequencies $\omega_{\text{THz}} = 0.5, 1.0$ and $3.0$ THz, respectively. Furthermore, by sweeping the cavity dissipation rate $\kappa$ and the coupling strength $\chi$, we demonstrate that high-concurrence entanglement persists over a wide range of cavity parameters, confirming the resilience of the mechanism against variation in fabrication tolerances. Our analysis reveals that the constraints on $\Omega_\text{max}$ and $\gamma$ represent the primary physical limitations on the maximal entanglement attainable. For the protocol to function, the primary dressing of the emitters via the carrier drives should reach Rabi frequencies $\Omega_{R,i}$ nearly resonant to the THz cavity frequency---the optimum detuning $\Delta$ should be in the range of only $\sim\text{GHz}$, as shown in Fig.~\ref{fig:variation_drive}(b)---. A cap on the driving strength $\Omega_\text{max}$ increases the relative contribution of the detunings $\Delta_i$ to these Rabi frequencies,  effectively placing a ceiling on the  mixing angles $\theta_i$. This, in turn, caps the magnitude of the collective dissipation, since the Purcell-enhanced collective decay rates $\Gamma_i$ scale directly with $\theta_i$. Consequently, if the available drive is insufficient to push the system into a high-dressed-cooperativity regime, the Liouvillian gap fails to achieve the condition $\lambda\gg\gamma$. In such cases, the entanglement generation becomes too sensitive to the spontaneous emission of the emitters, setting a ceiling on the achievable steady-state concurrence. 
The plots also indicate the parameter regimes in which the approximations made remain valid: the condition for the adiabatic elimination of the cavity, $\kappa \gg c_i s_i \chi$, is met in the regions above the blue lines, while the condition for the RWA of the dissipation, $\omega_\text{THz} \gg \kappa$, is met in the regions below the red lines. The overlap of these regions corresponds to the bad cavity limit, which provides a good description not only of a cavity but also other dissipative channels, such as waveguides. Furthermore, extending into the strong-coupling regime---beyond the scope of our analytical treatment---still yields high entanglement, confirming the effect is robust and persistent in different parameter regimes.

\subsection{Optical quantum state tomography}
We now introduce a method to perform quantum state tomography (QST) of the joint steady-state density matrix of the emitters $\hat\rho$--allowing to characterize the generated Bell state---employing only optical measurements and control [see Fig.~\ref{fig:tomography}(a-b)].
Since we have $N = 2$ emitters (each having $3$ coordinates labeled by $X,Y,Z$), QST requires to measure across $N_U = 9$ different joint coordinates. This enables complete state reconstruction through linear inversion, as illustrated in Fig.~\ref{fig:tomography}(b). In our setup, we can perform measurements of the $Z$ coordinate of each emitter---i.e. the population of the excited state--- by detecting the optical fluorescence using a ring-down protocol, whereby, once the steady-state has been reached, all the drives are turned off and the system is left to decay via spontaneous emission. We assume that each of the emitters are  spatially separated, allowing to resolve the fluorescence of each of them individually. To measure the $X$ and $Y$ coordinates, we apply local basis rotations to each of the emitters $i$ immediately prior to the ring-down fluorescence measurement. In general, access to the $(X,Y,Z)$ coordinates is achieved via  $\hat U_i\in\{\hat {\text H},\hat {\text S}^\dagger\hat{\text H},\hat I\}$ unitary rotations, implemented with coherent optical pulses. Here, $\hat{\text H}$ is the Hadamard gate, generated from a Hamiltonian $\hat H_X = -\Omega_U(\hat\sigma_x+\hat \sigma_z)/\sqrt{2}$, and $\hat{\text{S}}$ is the phase gate, with the rotation gate ${\text S}^\dagger\hat{\text H}$ being generated from a Hamiltonian $\hat H_Y = \Omega_U\hat \sigma_y/2$. Each of these Hamiltonians are obtained via optical drives with the appropriate frequency and phase, which are activated during a time $T = \frac{\pi}{2\Omega_U}$. To avoid decoherence during this step, the pulse duration must be significantly shorter than the spontaneous emission lifetime ($T\ll1/\gamma$). 
By setting $T=1/(10\gamma)$, we require a rotation Rabi frequency $\Omega_U = 5\pi\gamma$. For our chosen emitter linewidth of $\gamma/2\pi \approx 39.79$ MHz, this requires a rotation frequency of $\Omega_U/2\pi \approx 90$ MHz, which comfortably fall within the limits of optical pulse control. After performing these rotations and measuring the fluorescence, we obtain the probability $p_{b_1b_2|U}$ of observing a specific emission pattern (where $b_i\in\{e,g\}$ represent whether emitter $i$ was ``bright'' or ``dark'') for each rotation $U$. The full protocol to measure a joint coordinate---composed of state preparation, application of rotations, and detection via ring-down measurements---is illustrated in Fig.~\ref{fig:tomography}(c). 

We apply error mitigation techniques (see Appendix~\ref{sec:appendix_QST-error-mitigation}) to account for detection errors such as detection efficiency $\eta_e = p(e|e)$ (probability of correctly detecting light when the emitter is in the excited state) and dark count suppression efficiency $\eta_g = p(g|g)$ (probability of correctly detecting the absence of light when the emitter is in the ground state). This allows us to obtain the ideal mitigated probabilities $p_{b|U}^{\text{ideal}}$ from the measured probabilities $p_{b|U}$. The reconstructed density matrix $\bar\rho$ is obtained via the linear inversion map~\cite{hu2022}:
\begin{equation}
    \bar{\rho{}} = \frac{1}{N_U}\sum_U\sum_{b\in\{e,g\}^{\otimes{}N}}{p}_{b|U}^{\text{ideal}}\bigotimes_{i=1}^N(3U_i^\dagger{}\ket{b_i}\bra{b_i}U_i-I).
\end{equation}
\begin{figure}[hbt!]
    \centering
    \includegraphics[width=0.95\linewidth]{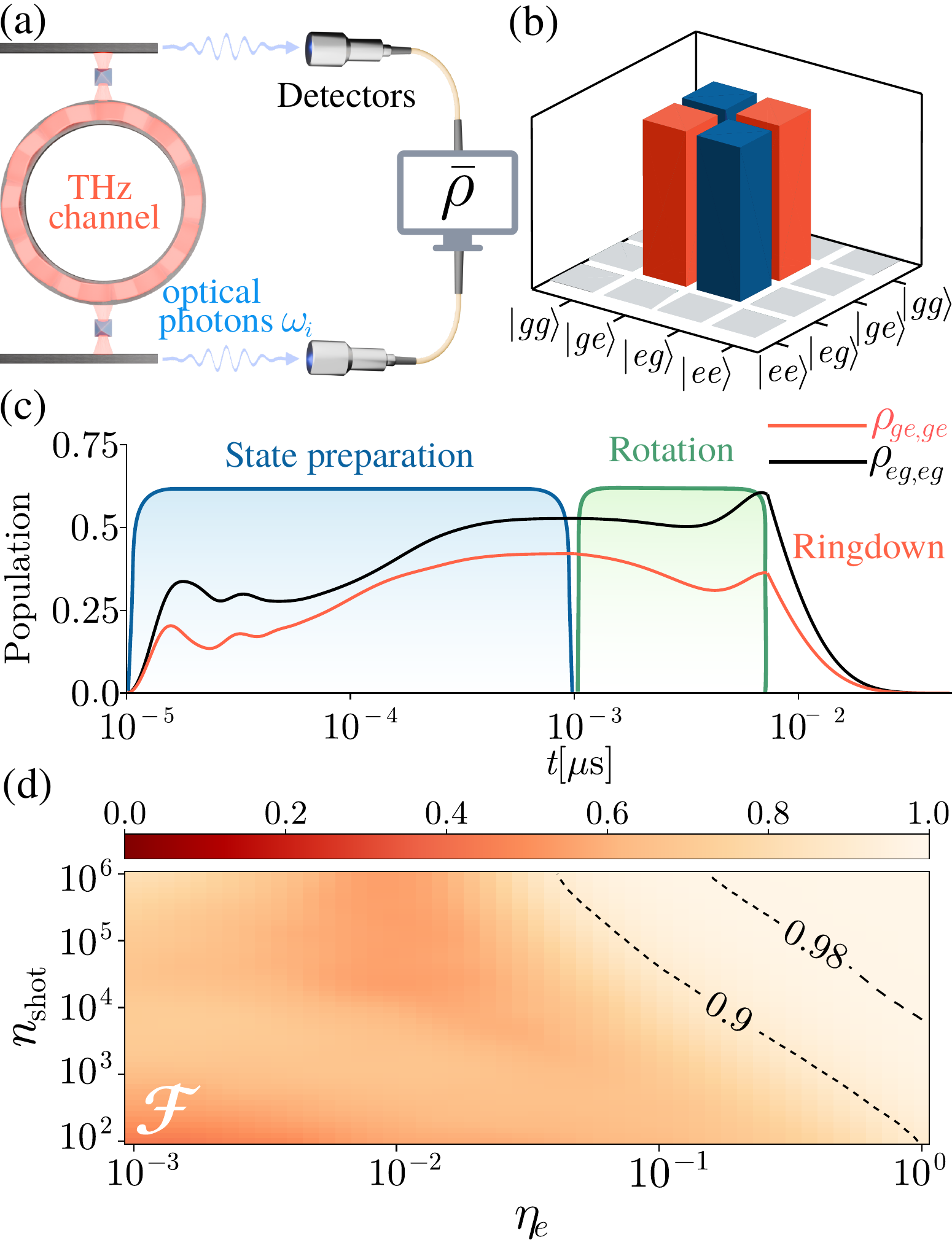}
    \caption{\textbf{Quantum State tomography (QST) and reconstruction Fidelity}. (a) Schematic of the QST acquisition protocol for two emitters radiating in the visible regime, based on joint photodetection (b) Reconstructed matrix for the maximally-entangled Bell state $\ket{\Phi^-} = \frac{\ket{ge}-\ket{eg}}{\sqrt{2}}$ using the linear inversion of Pauli measurement scheme. (c) Measurement protocol, consisting of dissipative entangled-state preparation, rotations $ U = H_1S_2^\dagger{}H_2$ and ringdown measurement, alongside the time evolution of relevant density matrix elements of the two-atom system. (d) Fidelity of the reconstructed density matrix relative to the generated density matrix in (c) after the state preparation, plotted as a function of the number of measurement samples  ($n_\text{shot}$) and the single-photon detection efficiency ($\eta_e$) with a fixed dark count loss $\eta_g = 0.99$. Each data point represents the mean value over $n_\text{ave}=50$ independent realizations. For visual clarity, a Gaussian smoothing with $\sigma=1.5$ has been applied to the resulting data. Parameters: (c) $\omega_{\text{THz}}/2\pi = 1.0$ THz, $\Omega_1/2\pi = 499.2$ GHz, $\Omega_2/2\pi = 493.1$ GHz, $\Delta_1/2\pi = 885.9$ GHz, $\Delta_2/2\pi = 874.9$ GHz, $\Omega_{\text{sb},1}/2\pi = 54.3$ GHz, $\Omega_{\text{sb},2}/2\pi = 55.4$ GHz,  $\chi/2\pi = 13.9$ GHz, $\kappa/2\pi = 25.6$ GHz, $\gamma/2\pi = 39.79$ MHz, $\Omega_U = \gamma$ and $T = \pi/(2\Omega_U)$.}
    \label{fig:tomography}
\end{figure}
Linear inversion preserves trace and hermicity of the reconstructed density matrix $\bar\rho$, but it does not guarantee it to be positive semi-definite~\cite{james2001,Kokaew2024}. To recover a valide physical state, further post-processing is required. While Maximum Likelihood Estimation (MLE) is widely used for optimal precision~\cite{Banaszek1999,Smolin2012}, it carries a significant computational cost. Here, we apply one of the simplest methods, consisting of removing eigenstates with negative eigenvalues from the eigendecomposition of $\bar\rho$~\cite{kaznady2009}. For highly entangled two-qubit states--- such as the high-concurrence targets in our protocol---this straightforward truncation method yields highly accurate reconstructions, providing a rapid and reliable verification of entanglement without the computational overhead of MLE~\cite{kaznady2009}. Figure~\ref{fig:tomography}(d) shows the fidelity of the resulting mitigated QST reconstruction as a function of the detection efficiency $\eta_e$ and the number of measurement shots $n_\text{shot}$. The results demonstrate that high fidelity reconstructions are feasible even under moderate efficiency. 
Non-monotonic dependence with $n_\text{shot}$ and $\eta_e$ are explained by divergences in the error mitigation protocol when $\eta_e = 1-\eta_g$ (see Appendix~\ref{sec:appendix_QST-error-mitigation}). 

From this fidelity map, we can evaluate the temporal cost of the procedure. Since the entangled state stabilization relies on Purcell decay rates larger than $\gamma$, the experimental duration is primarily limited by the photon collection period during the ring-down measurement [see Fig.~\ref{fig:tomography} (c)]. To allow for full signal decay, we estimate that the total time scales as $T\approx10n_\text{shot}\gamma^{-1}$. For a representative sample of $10^6$ shots the entire tomography process is completed in approximately $T \approx0.3$s, confirming the experimental feasibility of the verification protocol.

\section{Conclusion}
We have demonstrated a dissipative mechanism, mediated by a THz mode, that generates steady-state entanglement between polar emitters driven by visible light.
Our results predict robust entanglement across a broad spectral window of THz modes and over a wide range of experimentally accessible parameters. We show that both tuning and control can be achieved using purely optical knobs, such as laser detuning and drive amplitudes. Crucially, state detection is also entirely optical: steady-state cross-correlations between photons emitted by the two emitters provide a direct entanglement witness, while a complete framework for quantum state tomography enables full reconstruction of the entangled state via optical ring-down measurements.

Our scheme establishes a hybrid visible–THz interface, realizing a key operational primitive for THz quantum technologies, in which a THz channel mediates and stabilizes entanglement. The realization of these two-qubit protocols also opens the exploration of quantum networks with larger number of quantum emitters that exploit the unique spectral and suitable scaling properties of the THz domain. Although our analysis focuses on the THz regime, the protocol is general and applicable to hybrid architectures characterized by disparate energy scales, bridging the gap between vastly different frequency domains.

\section*{Code and Data Availability}
The data that support the findings of this study are available in \cite{lefur2026b_code}. The numerical simulations were performed using \verb|QuantumToolbox.jl|~\cite{mercurio2025} and \verb|Optimization.jl|~\cite{vaibhav_kumar_dixit_2023_7738525}.
\section*{Acknowledgments}
We thank Joan Agustí for useful insights and discussions. CSM and YLF acknowledge support by the project PID2023-149969NA-100 (SEQUOIA) funded by the Spanish Agencia Estatal de Investigación MICIU/AEI/10.13039/501100011033, and by a 2025
Leonardo Grant for Scientific Research and Cultural Creation from the BBVA Foundation. DMC acknowledges support from  MCINN projects PID2021-126964OB-I00 (QENIGMA), TED2021-130552B-C21 (ADIQUNANO) and PID2024-156077OB-I00 (DQUOTE). The authors acknowledge the support by the project NPhoQuss funded by Programa Fundamentos 2024 from the BBVA Foundation through the grant EIC24-1-17304.
\appendix{}
\section{GENERALIZED ROTATING WAVE APPROXIMATION}
\label{app:polaron}
In this appendix, we provide the analytical derivation of the generalized rotating wave approximation (GRWA) performed to obtain Eq.~\eqref{eq:ad_hamiltonian} and~\eqref{eq:Hc}. We take as a starting point the Hamiltonian presented in Eq.~\eqref{eq:dressed_hamiltonian}. The GRWA is performed in two step: first, a polaron transformation~\cite{debernardis2025,diaz-camacho2016,roman-roche2022} that absorbs the longitudinal couplings followed by a RWA that retains only the energy-conserving transitions in the frame rotating at cavity frequency $\omega_{\text{THz}}$. In our system, the polaron transformation is defined by the unitary $U_P = \exp\left[-\zeta(\hat{a}-\hat{a})(\hat{P}_1+\hat{P}_2)\right]$ where $\hat{P}_i = 1+(c_i^2-s_i^2)\hat{\xi}_z^i$. Applying this transformation with a specific choice of $\zeta = \chi/\omega_{\text{THz}}$ effectively eliminates the longitudinal coupling terms $\propto(\hat{a}+\hat{a})\hat{\xi}_{z,i}$ from  Eq.~\eqref{eq:dressed_hamiltonian}. This yields the transformed Hamiltonian:

\begin{multline}
        \tilde{H} =\omega_{\text{THz}}\hat{a}^\dagger{}\hat{a}+\frac{\Omega_{R,1}}{2}\hat\xi_{z,1}+\frac{\Omega_{R,2}}{2}\hat\xi_{z,2}-\zeta\chi(\hat P_1+\hat P_2)^2\\
        +\sum_{i=1,2} \biggl\{[4\chi\zeta(\hat P_1+\hat P_2)-2\chi(\hat{a}+\hat{a}^\dagger{})]
    c_is_i \, \hat\xi_{P,x,i} \\
    + \frac{\Omega_{\text{sb},i}}{2}[(2c_is_i\hat\xi_{z,i}+c_i^2\hat\xi_{P,i}-s_i^2\hat\xi_{P,i}^\dagger)e^{i\omega_{\text{THz}}{}t}+\text{H.c}]\biggr\},
    \label{eq:polaron_hamiltonian}
\end{multline}
where $\hat \xi_{P,i} = \hat \xi_ie^{-\zeta(c_i^2-s_i^2)(\hat a^\dagger{}-\hat a)}$ are the transformed dressed operators in the polaron frame. This transformation introduces a new nonlinear term, $-\zeta\chi(\hat P_1+\hat P_2)^2$. Expanding this term (and neglecting global energy shifts) yields an effective Lamb shift with rate $\Lambda_i \equiv 8\chi^2(c_i^2-s_i^2)/\omega_{\text{THz}}$, and an effective qubit-qubit interaction with rate $J \equiv 2\chi^2(c_1^2-s_1^2)(c_2^2-s_2^2)/\omega_{\text{THz}}$. Assuming a weak coupling regime relative to the cavity frequency ($\omega_{\text{THz}}\gg\chi$), we can expand the exponential operators $\hat \xi_{P,i}$ in Eq.~\eqref{eq:polaron_hamiltonian} to first order in $\zeta$, using $e^{\pm\zeta(s_i^2-c_i^2)(\hat a^\dagger{}-\hat a)} = 1\pm\zeta(s_i^2-c_i^2)(\hat a^\dagger{}-\hat a)+\mathcal{O}(\zeta^2)$. Substituting this linear expansion into the Hamiltonian in Eq.~\eqref{eq:polaron_hamiltonian} sets the stage for the RWA. To strictly justify the truncation of rapidly oscillating terms at THz frequencies, we must also assume that the sideband driving amplitude is sufficiently weak ($\Omega_{\text{sb},i}\ll\omega_\text{THz}$). Applying the RWA under these conditions yields the final GRWA Hamiltonian presented in Eq.~\eqref{eq:ad_hamiltonian} and~\eqref{eq:Hc}.

\section{CONDITIONS FOR A DARK STEADY STATE}
\counterwithin{figure}{section}
\label{app:doubly-dressed}
Here, we provide the necessary conditions to obtain a dark steady state from the dynamics described by the Hamiltonian in (\ref{eq:doubly_dressed_hamiltonian}) and jump operators in (\ref{eq:sideband_jumps}). For a state to be considered dark, it must be decoupled from the environment and dynamically stationary, in turn, the dark state must be annihilated by the jump operators $J_i$ such that $J_i\ket{D} = 0$ and an eigenstate of the Hamiltonian $H\ket{D} = E\ket{D}$~\cite{fleischhauer2000,zhou2022}. The condition on the jump operators impose to the doubly-dressed mixing factors that $\sqrt{\Gamma_1}\tilde{c}_1\tilde{s}_1 = \sqrt{\Gamma_2}\tilde{c}_2\tilde{s}_2$~\cite{govia2022}. In the symmetric case ($\Gamma_1 = \Gamma_2$), this simplifies to the complementary angle condition ($\tilde\theta_2 = \pi/2-\tilde\theta_1$). Then, in the absence of qubit-qubit interaction ($J = 0$), the Hamiltonian eigenstate condition can be fulfilled by matching Rabi frequencies ($\tilde\Omega_{R,1} = \tilde\Omega_{R,2}$). However, a finite qubit-qubit interaction requires a correction. When neglecting the counter-rotating terms of (\ref{eq:doubly_dressed_hamiltonian}), the solution of the eigenstate problem $H\ket{D} = E\ket{D}$ can be obtained by solving the set of equations:
\begin{equation}
\left\{
    \begin{split}
    & \frac{\tilde{\Omega}_{R,2}-\tilde{\Omega}_{R,1}}{2}-J\sin{}(2\tilde\theta_1)\sin{}(2\tilde\theta_2)\frac{\sqrt{\gamma_1}\tilde{c}_1^2}{\sqrt{\gamma_2}\tilde{c}_2^2}=E,\\
    & \frac{\tilde{\Omega}_{R,1}-\tilde{\Omega}_{R,2}}{2}-J\sin{}(2\tilde\theta_1)\sin{}(2\tilde\theta_2)\frac{\sqrt{\gamma_2}\tilde{c}_2^2}{\sqrt\gamma_1\tilde{c}_1^2}=E.
    \end{split}
    \right.
\end{equation}
The solution to this system of equations gives the condition in Eq.~\eqref{eq:condition-2}.

\section{LIOUVILLIAN GAP}
\label{app:Lgap}
The Liouvillian gap defines the speed at which the system converges to the steady-state but also the robustness of the system to decoherence. In this appendix, we give an analytical description to the Liouvillian gap $\lambda$. Taking as a starting point the dynamics presented in (\ref{eq:doubly_dressed_hamiltonian}) and (\ref{eq:sideband_jumps}). For no qubit-qubit interaction ($J = 0$), we have in the symmetric case ($\Gamma_1 = \Gamma_2$) that the Liouvillian gap is:
\begin{multline}
        \lambda = -\frac{9+3\cos(4\tilde\theta_1)}{8}\Gamma_1\\
        +\frac{\Gamma_1}{16}\sqrt{-40\cos(4\tilde\theta_1)+18\cos(8\tilde\theta_1)+86},
\end{multline} 
In the strong dressing regime, we have $\tilde\theta_1 \approx\pi/4-\frac{\Delta_R}{2\Omega_\text{sb}} $.
Expanding the expression for $\lambda$ up to second order we have $\lambda\approx -\frac{4\Gamma\Delta_R^2}{3\Omega_\text{sb}^2}$. 
While this expression is valid in the strong-driving limit, this will break when $\tilde\Omega_{R,i}\leq\Gamma_i$ and more general expression will be needed. This  can be obtained via the hierarchical adiabatic elimination, as described in Ref.~\cite{vivas-viana2025}.

\section{PARAMETER OPTIMIZATION}
\label{sec:appendix_optim_params}
Here we detail the steps taken to reduce the a six-dimensional parameter space --- consisting of the carrier and sideband drives $(\Omega_i, \Delta_i, \Omega_{\text{sb},i})$ --- down to two free parameters, $(\tilde\Omega_{R,1}, \tilde\theta_1)$ to be optimized.
To ensure experimental feasibility, we restrict the maximum Rabi frequency to $\Omega_\text{max}/2\pi = 0.5$ THz~\cite{boos2024}. The parameter reduction proceeds as follows:
\begin{enumerate}
    \item \textbf{Carrier reparametrization:} By expressing the carrier parameters in polar coordinates as $(\Omega_i,\Delta_i)= (\Omega_{R,i}\sin2\theta_i,\Omega_{R,i}\cos2\theta_i)$, we map the free variables to $(\Omega_{R,i},\theta_i)$.
    \item \textbf{Applying Condition 0:} This condition [see Eq.~\eqref{eq:condition_0}] enforces equal primary-dressing angles, ${\theta_1 = \theta_2}$, removing one free parameter (5 left). Because the protocol benefits from maximizing the collective decay $\Gamma_i$, we lock one of the carrier drives to the maximum allowed value, $\Omega_1 = \Omega_\text{max}$ (we cannot lock both since we need $\Omega_{R,1}\neq \Omega_{R_2}$).  This removes another free parameter (4 left), and fixes the angles to $\theta_1 = \theta_2 =  \theta_\text{max} = \frac{1}{2}\arcsin(\Omega_\text{max}/\Omega_{R,1})$. Here we assumed that $\Omega_{R,2}\leq\Omega_{R,1}$, implying  $\Omega_2\leq\Omega_\text{max}$. This leaves four free parameters $(\Omega_{R,i},\Omega_{\text{sb},i})$, which are equivalent to $(\Delta_{R,i},\Omega_{\text{sb},i})$ [see Eq.~\eqref{eq:ad_hamiltonian}].
    \item \textbf{Applying Conditions 1 and 2:} Analogously to previous steps, we reparametrize $(\Delta_{R,i},\Omega_{\text{sb},i})\rightarrow (\tilde\Omega_{R,i},\tilde\theta_i)$. Finally,  imposing Condition 1 (complementary angles),  $\tilde\theta_2 = \pi/2-\tilde\theta_1$,  and Condition 2 (overlapping sidebands), $\tilde\Omega_{R,2} = \tilde{\Omega}_{R,1}$, constrains the system to exactly two independent variables: the doubly-dressed Rabi frequency  $\tilde\Omega_{R,1}$ and the mixing angle $\tilde\theta_1$. 
\end{enumerate}

By working with qubit-laser parameters written in the doubly-dressed basis $(\tilde\theta, \tilde \Omega_R)$ we can straightforwardly apply Conditions 1-2 and solve the ME in Eq.~\eqref{eq:doubly_dressed_hamiltonian} and \eqref{eq:sideband_jumps}, provided we also have established the values of $\Gamma$ and $J$, that depend on the only remaining qubit-laser parameter to fix, the dressed angle $\theta$ of the primary-dressed basis. This angle $\theta$ appears as $c=\cos\theta$ and $s=\sin\theta$ in the expressions for $J$ and $\Gamma$. 
For a given $\tilde \Omega_R$, $\tilde\theta$ and $\Omega_\text{max}$, we can recover $c$ and $s$. The expression to solve is
\begin{equation}
    \sin2\theta = \Omega_\text{max}/\Omega_{R,1}.
    \label{eq:theta-OmegaR1}
\end{equation}
The value of $\Omega_{R,1}$ is retrieved from the set of doubly-dressed parameters fixed earlier.
In particular
\begin{equation}
   \tilde{\Omega}_{R,i} = \Delta_{R,i}/\cos2\tilde\theta.
\end{equation}
Writing explicitly $\Delta_{R,i}$ we get
\begin{equation}
   \tilde{\Omega}_{R,1} = ( \Omega_{R,1}-\omega_{\text{THz}}- 8\chi^2(c_1^2-s_1^2)/\omega_{\text{THz}} )/\cos2\tilde\theta.
\end{equation}
Solving for $\Omega_{R,1}$ and substituting in Eq.~\eqref{eq:theta-OmegaR1} gives an exact but cumbersome expression of $c^2$. We have checked that, for the range of values of $\chi^2/\omega_\text{THz}$ considered in this work, we can safely assume
\begin{equation}
   \tilde{\Omega}_{R,1} \approx ( \Omega_{R,1}-\omega_{\text{THz}}- 8\chi^2/\omega_{\text{THz}} )/\cos2\tilde\theta,
\end{equation}
which provides simple analytical expressions for $\theta$ that only yield a relative error from the optimal values of $J$ and $\Gamma$ up to the third significant digit.
Finally, we show in Fig.~\ref{fig:optimal_omega_r} and Fig.~\ref{fig:optimal_theta_1} the optimal parameters ($\tilde{\Omega}_{R,1},\tilde\theta_1$) found to maximize the stationary concurrence presented in Fig.~\ref{fig:optimization_map}.
\begin{figure}[h!]
    \centering    \includegraphics[width=0.95\linewidth]{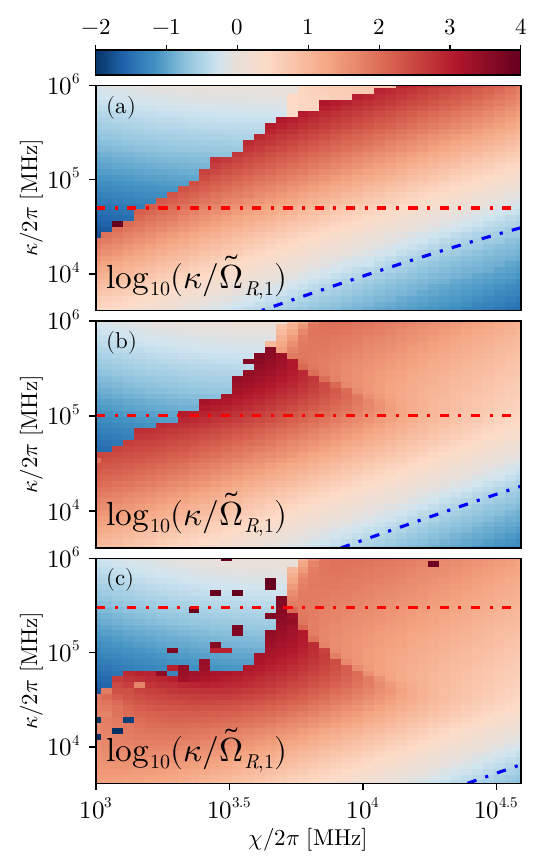}
    \caption{Optimal values of the Rabi frequency $\log_{10}(\kappa/\tilde{\Omega}_{R,1})$ required to achieve the maximal stationary concurrence $\mathcal{C}$ in Fig.~\ref{fig:optimization_map} versus the cavity parameters ($\chi$,$\kappa$) for (a) $\omega_\text{THz} = 0.5 $ THz (b) $\omega_\text{THz} = 1.0 $ THz and (c) $\omega_\text{THz} = 3.0 $ THz. The dot dashed lines represent the validity of the RWA (in Red) and the adiabatic elimination (in blue).}
    \label{fig:optimal_omega_r}
\end{figure}
 
\begin{figure}[h!]
    \centering
    \includegraphics[width=0.95\linewidth]{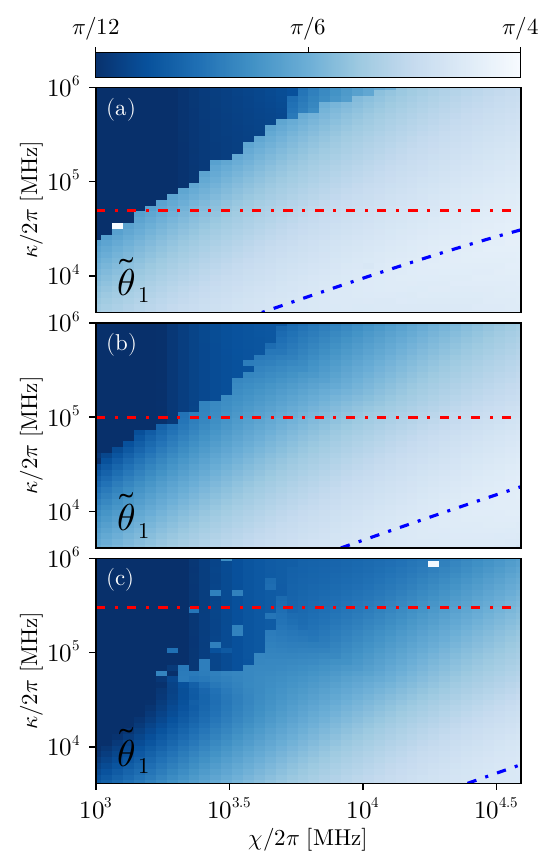}
    \caption{Optimal values of the doubly dressed angle $\tilde\theta_1$ required to achieve the maximal stationary concurrence $\mathcal{C}$ in Fig.~\ref{fig:optimization_map} versus the cavity parameters ($\chi$,$\kappa$) for (a) $\omega_\text{THz} = 0.5 $ THz (b) $\omega_\text{THz} = 1.0 $ THz and (c) $\omega_\text{THz} = 3.0 $ THz. The dot dashed lines represent the validity of the RWA (in Red) and the adiabatic elimination (in blue).}
    \label{fig:optimal_theta_1}
\end{figure}

\section{ERROR MITIGATION FOR STATE TOMOGRAPHY}
\label{sec:appendix_QST-error-mitigation}

We take into account  detection errors such as detection efficiency $\eta_e = p(e|e)$ (probability of correctly detecting light when the emitter is in the excited state) and dark count suppression efficiency $\eta_g = p(g|g)$ (probability of correctly detecting the absence of light when the emitter is in the ground state).

The relationship between our measured probabilities ($p_U$) and the ideal noise-free probabilities ($p_U^{\text{ideal}}$) can be described by a confusion matrix $\Lambda$~\cite{maciejewski2020,bravyi2021,hashim2025}} as $p_U = \Lambda p_U^{\text{ideal}}$. Assuming identical detector performance, this is given by 
\begin{equation}
    \Lambda = \bigotimes_{i=1}^N\begin{pmatrix}
        \eta_g&1-\eta_e\\
        1-\eta_g&\eta_e
    \end{pmatrix}
\end{equation}
In this matrix, the off-diagonal terms represent the misidentification events. We can mitigate these detection errors and recover the ideal probabilities  ($p_U^{\text{ideal}}$) simply by inverting the confusion matrix~\cite{maciejewski2020,bravyi2021}. 
However, simple matrix inversion is highly sensitive to statistical fluctuations. Because the inverse matrix scales with $1/\det\Lambda$, where for a single emitter we get $\det\Lambda = \eta_e+\eta_g-1$, any noise due to the finite shot noise is drastically amplified as the system approach the singular point where $\eta_e = 1-\eta_g$. At this point, the determinant vanishes ($\det\Lambda = 0$) making the direct inversion impossible~\cite{maciejewski2020,nachman2020}. In Fig.~\ref{fig:tomography}(d),  the resulting degradation of the averaged fidelity is noticeable at $\eta_e = 1-\eta_g = 0.01$.

In a laboratory setting, if an experiment lands precisely on this singularity, one could bypass the inversion undefiniteness by intentionally degrading one of the detector efficiencies ( for instance artificially lowering $\eta_e$) to break the symmetry and force a non-zero determinant. While this restore the invertibility of $\Lambda$, the inversion mitigation can still lead to variance amplification. To robustly circumvent this limitation one can apply more optimized mitigation protocols such as SVD unfolding~\cite{nachman2020} or Maximum Likelihood Estimation (MLE)~\cite{maciejewski2020}.

\let\oldaddcontentsline\addcontentsline
\renewcommand{\addcontentsline}[3]{}
\bibliography{THzEntanglement}
\let\addcontentsline\oldaddcontentsline
\end{document}